# Tuning the average path length of complex networks and its influence to the emergent dynamics of the majority-rule model

Andreas I. Reppas[*], Konstantinos Spiliotis, Constantinos I. Siettos[†]

School of Applied Mathematics and Physical Sciences, National Technical University of Athens, Greece, GR 157 80

**Abstract.** We show how appropriate rewiring with the aid of Metropolis Monte Carlo computational experiments can be exploited to create network topologies possessing prescribed values of the average path length (APL) while keeping the same connectivity degree and clustering coefficient distributions. Using the proposed rewiring rules, we illustrate how the emergent dynamics of the celebrated majority-rule model are shaped by the distinct impact of the APL attesting the need for developing efficient algorithms for tuning such network characteristics.

**1. Introduction.**

The strong interplay between the emergent complex dynamics of many real-world systems and the underlying topology of the networks that pertain to their structure has been illustrated by many studies: the dynamics of electrical power transmission systems including cascade failures leading to blackouts [54], the evolution of the world wide web, various social phenomena such as mimesis and herding [55], brain cognitive, neurological disorders and motor functions [8, 11, 54] are typical paradigms of such cases.

Thus, the modelling and the systematic investigation of the topological properties of complex networks are of great importance. Towards this aim, various algorithms for generating networks aspiring to approximate the actual ones have been proposed [2, 46, 57].

In order to approximate such real-world network structures, Watts and Strogatz [57] (WS) constructed a network model with a variable connectivity degree possessing "small-world" properties interpolating between regular ring lattices and random regular networks (RRN). Small-world networks are highly clustered with small path lengths. The most famous experiment which describes the concept of "small-worldness" was conducted by the physiologist Milgram [44]. He sent 160 letters to residents of the city Omaha in Nebraska asking them to post a letter to a friend.

---

[*] *Currently, Technische Universität Dresden, Center for Advancing Electronics, Dresden, Germany*

[†] Corresponding author Tel.: +30-210772-3950, e-mail adress: ksiet@mail.ntua.gr

The shipment had only one term: "do not try to send the mail directly if you don't know the recipient; instead, post the mail to one friend of yours who you believe knows the recipient". The letters, which were finally received, had been posted about six times on average. This phenomenon is also known as six degrees of separation (also known as "six degrees of Kevin Bacon" [57]). Other networks, such as the Internet, include nodes with overwhelmingly more connections compared to other nodes (for instance yahoo, google and amazon are nodes with a huge number of connections). This type of networks is usually characterized by a power-law distribution of degrees i.e. $P(k) = ck^{-\gamma}$ and are called scale-free. Barabasi and Albert [6] in their seminal paper proposed an algorithm for generating networks with power-law degree distributions. These networks can capture the power-law characteristic which pertains to the structure of many real-world networks, yet the clustering coefficient decays very fast with network size; therefore, failing to approximate larger clusters as these are observed in many real life structures [35]. Furthermore, there are many social networks whose structure deviates from the power-law degree distribution or small-worldness (most often exhibiting skewed degree distributions [12]). In addition, the topological characteristics of many networks may change over time. For example in [49], it is shown that in the contact social network of a disease transmission at an American high school the clustering coefficient remained almost constant over a wide range of contact durations while the average path length doubles.

In order to generate networks structures that can capture the empirically observed ones, research efforts have been focused on developing algorithms for generating network topologies with prescribed characteristics. In [28], the authors propose a network-growing algorithm combining the properties of both scale-free networks and small-world networks. The value of the clustering coefficient is driven by manipulating the formation of triades. Serrano and Boguna [50] present a network-growing algorithm for controlling both the degree distribution and the clustering coefficient. Their algorithm is based on the so-called configuration model which is used to generate pre-assigned degree distributions. The most typical representative of this category is the celebrated Erdòs-Rényi algorithm [20]. Volz [56] uses a Markov chain Monte Carlo technique to generate both a given degree distribution and a clustering coefficient. Maslov and Sneppen [41] use a rewiring technique to produce random networks, with a given connectivity degree, in the case of the interactions of nuclear proteins. Kim [33] introduces an algorithm based on a Monte Carlo simulation at both zero and finite temperatures to control the clustering coefficient of a given network. In [24], the authors propose an algorithm for generating small-world networks with tunable assortative coefficient. Leary et al. [38] present an algorithm for controlling the degree distribution by altering the preferential attachment step in the Barabàsi and Albert algorithm. Exploiting the algorithm of Holme and Kim, they were able to produce different degree distributions with different clustering coefficients. Badham and Stocker [5] propose an algorithm for adjusting three properties of networks, namely the degree distribution, the clustering coefficient and the assortativity. In [21], the authors propose an optimization method for constructing a network with prescribed degree-dependent clustering.

In this work, we propose appropriately chosen rewiring rules that can be used to systematically construct consistent to APL network structures at will, yet maintaining the degree and clustering distributions untouched. To demonstrate the approach, we constructed networks (starting from small-world networks which served as our initial configurations derived using the WS algorithm [57]), with prescribed values of APLs. At this point we should note that upon the construction of a small-world or a scale-free network using an algorithm such as the one proposed in [57] or in [2], the APL cannot be in principle prescribed.

Furthermore, we show how different topologies as these are obtained by adjusting the value of the APL can dramatically shape the emergent dynamics of network-based models. For our illustrations, we chose a significant representative of such cases: the majority-rule model. Majority-rule models have been extensively used to simulate and gain a better understanding of the behaviour of many complex systems ranging from epidemic spread dynamics [9,10, 31, 42] and opinion formation and voter/election dynamics [14, 27, 37, 53] to culture and language dynamics [4, 16, 18, 19, 40], crowd flow design and management [25, 26, 29, 48], diffusion of news and innovations [23, 39, 58] ecology and neuroscience [36, 51].

## 2. Tuning the Average Path Length of a Complex Network.

The clustering coefficient and the APL of a network are two attributes that contain significant information concerning its topological structure. The APL, say $L$, is a global property indicating the average number of steps required to reach any two nodes. It is defined as the mean value of all shortest paths between any two nodes, i.e. $L = \dfrac{\sum d_{i \leftrightarrow j}}{\dfrac{N(N-1)}{2}}$, where $d_{i \leftrightarrow j}$ is the shortest path between the nodes $i$ and $j$ and $N$ is the number of nodes in the network. On the other hand, the clustering coefficient expresses a "local" characteristic of a node regarding the formation of cliques among its neighbors. In other words, as in the case of social networks, it measures the fact that "friends of my friends are also likely to be friends with each others" [46]. The clustering coefficient $c_i$ of a node $i$ is defined as follows: let $k_i$ be the degree of node $i$, i.e. the number of edges connected to node $i$. If the number of possible edges between the $k_i$ neighbours (or the total number of possible triangles) of node $i$ is $\dfrac{k_i(k_i-1)}{2}$ and the number of edges that really exist is $E_i$ (i.e. the number of existing triangles), then the clustering coefficient $c_i$ is defined as: $c_i = \dfrac{2E_i}{k_i(k_i-1)}$. The clustering coefficient, say $C$, of the whole network is defined as the mean value of the clustering coefficients $c_i$ of every node, i.e. $C = \dfrac{1}{N}\sum_{i=1}^{N} c_i$, where $N$ is the number of nodes in the network.

Here, we present an algorithm, which can be used to change the structure of a certain network by tuning the APL without altering the underlying degree distribution and the clustering coefficient of any node, i.e. keeping the clustering distribution untouched. The approach is based on appropriately chosen rewiring and generally can be applied to (i) simple networks, where multiple edges or self-loops among the nodes are not permitted, (ii) undirected nodes, where if node $i$ has node $j$ as a neighbor, then $j$ has $i$ as a neighbor and (iii) one-component networks, in which starting from a node, one can find a path to visit all the other nodes. Although this last condition is not vital for the rewiring process and can be excluded, it ensures the calculation of the average path length over the whole network.

To obtain the desired values for the average path length of a given network, the proposed algorithm combines appropriate rewiring with Simulated Annealing (SA) [1, 13, 34]. For the APL, the objective function at the step $k$ of the SA algorithm may be defined as: $E^k = \left\| L^k - L^{\text{target}} \right\|$, where $L^{\text{target}}$ is the target value. The proposed algorithm can be summarized as follows (see also Figure 1):

*Step 0) Set the initial system's pseudo-temperature, say Temp and select the annealing scheme, i.e. the way the pseudo-temperature will decrease.*

*Do until convergence {*

*Step 1) Evaluate the pseudo-energy (objective function) $E(L) = \left\| L - L^{\text{target}} \right\|$ of the network.*

*Implement the following rewiring rules {*
*Step 2) Select randomly two nodes $i$ and $j$ which*
  *(2a) do not have any common neighbors and*
  *(2b) each of them has at least one neighbor (say, $i_1$ and $j_1$ respectively) that does not form any triangle with any other neighbor of $i$ and $j$ nor do any common neighbors exist between $i_1$ and $j_1$ (see Figure 1a).*

*Step 3) Rewire the edges to connect $i$ with $j$ and $i_1$ with $j_1$ (Figure 1b). At this point the rewiring process may produce multiple components in the network and if it does the new configuration should be rejected and the algorithm returns to step 2.*
*}*
*Step 4) Evaluate the new mean path length of the network, say $L'$ and the corresponding objective function $E(L')$.*

*Step 5) Accept or reject the new configuration using the Metropolis procedure [48]:*
  *5a) Accept the new configuration if $E(L') < E(L)$,*
  *5b) Accept the new configuration if $E(L') > E(L)$ with a probability $\exp\left[-\left(E(L') - E(L)\right)/Temp\right]$ ;otherwise reject it.*

*5) Reduce the system pseudo-temperature according to the annealing schedule.*

*} End do.*

In order to test the efficiency of the proposed scheme, we initially constructed "small-world" networks using the Watts and Strogatz (WS) algorithm [57]: starting from a ring structure, i.e. a one-dimensional lattice of $N$ nodes, each of them connected with its $k$ nearest neighbours on both sides (i.e. each node has a connectivity degree of $2k$), links are rewired with a probability $p$. For a ring lattice (i.e. when $p = 0$), $L(N, p)$ scales as $L(N,0) \approx \frac{N}{4k} \sim N$ and its structure is characterized by a flat distribution of distances (shortest path lengths) between nodes (Figure 2). For completely random graphs (i.e. when $p \to 1$), the average path length scales as $L(N, p) \sim \frac{\ln(N)}{\ln(2k)}$ and the resulting structure is characterized by a very narrow distribution of shortest path lengths. For intermediate values of $p$ it has been shown [7, 45, 57], that there is a certain regime in which networks exhibit small-world properties characterized by narrowed-peaked distributions of shortest path lengths (Figure 2).

In our simulations, the initial system's pseudo-temperature was set to *Temp*=10, while the pseudo-temperature decreased 10% every 200 steps.

Here, we used two different network sizes, namely, $N = 1000$ and $N = 10000$, as well as different initial connections $2k = 6$ and $2k = 8$. When the probability is $p = 0$, there are no edges to be rewired without destroying the clustering formation of the network (step 2 of the algorithm), thus, the proposed algorithm cannot be implemented (Figure 3a). For values of the probability $0 < p \leq 1$ the proposed algorithm can find edges to rewire without changing the clustering coefficient and the connectivity degree of any node (Figure 3b).

In figure 4a, we present the implementation of the proposed approach starting from a network with $N = 1000$ nodes, $2k = 8$ constructed using the WS algorithm with a probability $p = 0.01$ of rewiring edges. We tried to adjust the value of the APL to lower as well as to higher values with respect to the WS network's values. As it can be seen, the algorithm reaches a maximum and minimum plateau. The proximity of the minimum plateau to the initial value of the APL is due to the fact that there are not many choices left in changing long-range interactions in order to lower the APL which at the same time leaves the degree and clustering distribution untouched. On the other hand, there are more choices for generating, by rewiring, connections that drive the APL in significantly higher values. Figure 4b illustrates the evolution of the probability distribution of the distance $d_{ij}$ between two nodes $i$ and $j$ as the average path length of the network increases. For comparison purposes, we also show the flat probability distribution of the ring network. It is obvious that, as the value of the APL gets higher, the resulting distribution of shortest path lengths gets flatter deviating as expected from the small-worldness.

Figure 5 depicts the APL (open circles) and clustering coefficient (open squares) of "small-world" networks with $N = 1000$ nodes, $2k = 6$, along with the maximum and minimum values of the average path lengths that the algorithm could reach for networks initially constructed with probabilities $p = \{0.005, 0.01, 0.2, 0.5, 0.9\}$ of rewiring edges. All other intermediate values of the APL could be successfully reached by the algorithm. It is worth mentioning that, while the networks constructed with probabilities $p \geq 0.2$ have an APL of $L(N,p)/L(N,0) \leq 0.06$, meaning that $L(N,p) \sim \ln(N)$, the algorithm succeeds to increase it up to $L/L(N,0) \geq 0.45$, meaning that $L(N,p) \sim N$. Hence the algorithm can create networks ranging from "small-world" to "large-world" structures while keeping the clustering and degree distributions constant.

Figure 6(a) presents the implementation of the algorithm in the case of a network with $N = 10000$ nodes, $2k = 6$ initially constructed with the WS algorithm with a probability $p = 0.5$ of rewiring edges. Figure 6(b) depicts the evolution of the probability distribution of $d_{ij}$ between two nodes $i$ and $j$. As in the previous case, when $N = 1000$, the demand for higher values of the APL of the network results to the generation of longer distances and flatter distributions of shortest paths.

The above simulations were obtained using an Intel(R) Core(TM) i5-3320M CPU @ 2.60 GHz CPU and 16 GB installed memory (RAM). Indicatively we note that for an increment of 10% of the APL of a network with N=10000 nodes and 30000 edges, the computational time was 12 hours on average. In the case of networks initially constructed with a rewiring probability p=0.25 the required computational time was approximately 150 hours (for the construction of 50 network configurations). In the case of the networks initially constructed with a rewiring probability p=0.5 the required computational time was approximately 187 hours (for the construction of 50 network configurations).

**3. The majority-rule model.**

Let us assume $N$ individuals interacting on a complex network. Each individual is labelled as $i$ ($i = 1,2,...,N$), and its state takes two values: one (denoting activation) and zero (denoting deactivation). Hence, the state of the $i^{th}$ in time $t$ can be described by the function $a_i(t) \in \{0,1\}$ [43]. Let us denote by $\Lambda(i)$ the set of the neighbors (i.e. the individuals connected to $i^{th}$ individual, with self loop included). Let us also consider the summation $\sigma_i(t) = \sum_{j \in \Lambda(i)} a_i(t)$, which gives the number of active neighbors of the $i^{th}$ individual. Then at each time step, each individual interacts with its neighbors and changes its state-value according to the following stochastic model [36, 51]:

1. An inactive individual becomes active with probability $\varepsilon$, if $\sigma_i(t) \leq \left(\frac{k_i+1}{2}\right)$ (where $k_i$ is the degree of the $i^{th}$ individual) and at least one of its links is active. If $\sigma_i(t) > \left(\frac{k_i+1}{2}\right)$, the individual becomes active with probability $1-\varepsilon$.

2. An activate individual becomes inactive with probability $\varepsilon$, if the $\sigma_i(t) > \left(\frac{k_i+1}{2}\right)$. If $\sigma_i(t) \leq \left(\frac{k_i+1}{2}\right)$, the individual becomes inactive with probability $1-\varepsilon$, where the range of the parameter $\varepsilon$ is the interval $(0, 0.5)$.

**4. Simulation Results.**

For our simulations we used networks of $N = 10000$ individuals. The initial one-dimensional ring was constructed with $2k = 6$ neighbours for every node and small world structures were constructed using the WS algorithm for values of the rewiring probability $p = 0.25, 0.7$. The parameter $\varepsilon$ of the majority rule model was set equal to 0.1. In each case, we used the proposed algorithm to test the "distinct" impact of the APL on the emergent dynamics of the majority-rule model by adjusting its value at prescribed values.

Figure 7(a) depicts the time evolution of the density of active individuals, say $d$, in the case of a small-world network constructed using the WS algorithm with a probability $p = 0.25$. The resulting network exhibits high clustering equal to $C = 0.2603$ and an APL equal to $L = 6.5984$ (averaged over 50 network configurations). As it is seen, for the specific value of the APL there are two stable stationary states, one corresponding to an "all-off" state, in which all the individuals are inactive, and the other one to a partially active network. Figure 7(b) depicts the time evolution of the density of active individuals when the APL was set to a slightly different value $L = 6.702$ while keeping the degree and clustering distribution the same. In this case, the second stationary solution loses stability and the network converges to the "all-off" state which is the only possible one. The temporal simulations dictate that there is a critical value of the APL that marks this phase transition. Hence, a slight change in the APL of the network influences significantly the emergent dynamics of the model.

Figure 8(a) shows the time evolution of the density of active individuals, $d$, in the case of a small-world network constructed using the WS algorithm with a probability $p = 0.7$ of rewiring edges. In this case the network has a low clustering coefficient $C = 0.0168$, and an APL equal to $L = 5.5149$ (averaged over 50 network configurations). As in the previous case, there are two stable stationary states, one corresponding to the "all-off" state and the other to a partially activated network. Implementing the proposed algorithm and increasing the APL to $L = 6.1588$, the non-zero steady state loses stability and the network converges to the "all-off" state revealing again a "critical" value for the APL (Figure 8(b)).

In Figure 9 we present the coarse-grained bifurcation diagram of $d$ with respect to the APL. The diagram was obtained by brute-force temporal simulations averaging over 50 different consistent realizations of networks. The resulting networks with the prescribed APL were created starting from small-world networks constructed with an initial probability $p = 0.25$ of rewiring edges. A critical threshold was found around $L_{cr} \cong 6.69$. The inset of Figure 9 depicts the stable stationary state which corresponds to the non-zero steady state. Figure 10 depicts the coarse-grained bifurcation diagram of $d$ with respect to the APL as obtained over 50 consistent network realizations. Here the resulting networks with the prescribed APL were created starting from small-world networks constructed with an initial probability $p = 0.7$ of rewiring edges. A critical threshold was found around $L_{cr} \cong 6.13$; the inset figure depicts the stable stationary state which corresponds to the partially activated network.

## 5. Conclusions and Discussion

Networks play an important role in many scientific areas ranging from chemistry to ecology and from biology to social sciences. The investigation of the interplay between topology and emergent dynamics of complex systems has been the focus of many studies. Through these studies has been made clear that the network topology, on which the interaction of the individuals/particles evolves, can shape the emergent macroscopic dynamics. However, it is less clear how one can quantify in a systematic manner the dependence of the emergent dynamics with respect to specific network characteristics. Due to the nonlinear, stochastic nature of such models and their coupling to complex network structures, the emergent behavior cannot be-most of the times-accurately modeled and analyzed in a straightforward manner.

Towards this direction, various algorithms for generating complex networks with specific topological characteristics ranging from completely random graphs with homogeneous degree distributions to small-world and scale-free structures, have been proposed. Furthermore, over the last years researchers have focused their efforts on developing algorithms for prescribing certain topological network properties such as the degree distribution, clustering and assortativity aspiring to approximate real-world observed networking.

Here we presented a framework that builds on earlier work on combining Monte Carlo simulations with rewiring that can be used to adjust at will the value of the average path length of a given network. The proposed rewiring rules and scheme allow the "at will" tuning of the APL leaving at the same time the degree and clustering coefficient distribution unchanged. For illustrative purposes, we created topologies with prescribed values of the APL coefficients starting from small world networks constructed using the Watts and Strogatz algorithm.

Exploiting the algorithm we were able to investigate the impact of the APL to the emergent dynamics of the celebrated majority-rule individualistic model pertaining to the dynamics of many real-world responses including herding under panic [3], the emergence of cooperation [47] and public opinion formation [30]. Our analysis revealed that even small changes in the APL (while keeping the other two important statistical topological characteristics, namely the degree and the clustering distributions untouched) can result in big changes in the system's behaviour. To our knowledge this is the first time that such an analysis is provided and demonstrates the scope of the tasks that one can attempt using the proposed framework, and how it may be used to draw more general conclusions about the distinct impact of this topological characteristic.

This interplay of the APL and majority-rule dynamics is of particular interest in processes like the spread of epidemics and information exchange, as one can ultimately use such an analysis to effectively control or stabilizes the spread rate (see for example [15]). Another aspect that it is worth pointing out is that the average path length is also directly associated to the so called weak links whose significance in many problems ranging from social to biological phenomena has been raised by many studies [17, 22]. Hence the proposed algorithm can be also used to shed more light on the influence of the weak links on the evolution of complex systems evolving on heterogeneous networks. At this point we should note that while our algorithm keeps the degree and clustering distributions untouched we should expect that other properties such as the assortativity and higher order motifs will generally change.

By employing the proposed algorithm, we constructed the bifurcation diagrams with brute-force temporal simulations. However, this is but the first task one would use to study the influence of the topology on the emergent dynamics: i.e. set up many initial network ensembles possessing the desired topological characteristics (here the APL); for each initial topology create a large enough number of consistent ensemble state realizations, and then run the detailed dynamics for a long time to investigate the system's behaviour. Yet, this simple simulation is inadequate for the systematic analysis and the construction of the complete bifurcating diagrams (for example unstable solution branches cannot be traced in this way). Due to the complex interplay between topology and emergent model dynamics, and the intrinsic multiplicity of scales at which the relevant individuals/ objects interact, the systematic analysis at the macroscopic/emergent level becomes an overwhelmingly difficult task. Usually, good macroscopic evolution equations in closed form cannot or it is difficult to be written in a straightforward manner. This limits our ability to analyse the emergent dynamics using well established numerical bifurcation analysis tools. When this is the case, one can exploit the potential of the Equation-Free approach [32] which serves as an on-demand identification-based approach enabling individual-based stochastic models to perform system-level tasks bypassing the need of deriving explicit models in a closed form. In [52] Spiliotis and Siettos showed how this multi-scale framework can be used to construct bifurcation diagrams and perform rare-events analysis with respect to the degree distribution of the underlying networks. An equivalent procedure can be applied in order to couple the proposed algorithm for the generation of network topologies with prescribed values of the APL with the Equation-Free approach in order to construct the complete bifurcation diagrams and perform systematic stability analysis.

**References**

[1] E. Aarts, J. Korst, W. Michiels, in: E.K. Burke, G. Kendall (Eds.), Search Methodologies: Introductory Tutorials in Optimization and Decision Support Techniques, second ed., Springer-Verlag, Berlin, 2006, pp. 187-210.

[2] R. Albert, A.L. Barabási, Statistical Mechanics of Complex Networks, Rev. Mod. Phys. 74, (2002) 47-97.

[3] E. Altshuler, O. Ramos, Y. Núñez, J. Fernández, A.J. Batista-Leyva, C. Noda, Symmetry Breaking in Escaping Ants, The American Naturalist 166(6) (2005) 643-649.

[4] R. Axelrod, The Complexity of Cooperation: Agent-Based Models of Competition and Collaboration, Princeton University Press, Princeton, 1997.

[5] J. Badham, R. Stocker, A Spatial Approach to Network Generation for Three Properties: Degree Distribution, Clustering Coefficient and Degree Assortativity, Journal of Artificial Societies and Social Simulation 13(1) (2010) 11-23.

[6] A.L. Barabási, R. Albert, H. Jeong, Mean-field theory for scale-free random networks, Physica A 272 (1999) 173-187.

[7] A. Barat, M. Weigt, On the properties of small-world network models, The European Physical Journal B 13(3) (1999) 547-560.

[8] D.S. Bassett, A. Meyer-Lindenberg, S. Achard, T. Duke, E. Bullmore, Adaptive of fractal small-world human brain functional networks, Proc Natl Acad Sci USA 103 (2006) 19518-19523.

[9] M. Boguñá, R. Pastor-Satorras, Epidemic Spreading in correlated complex networks, Phys. Rev. E 66 (2002) 047104.

[10] M. Boots, A. Sasaki, 'Small worlds' and the evolution of virulence: infection occurs locally and at a distance, Proc Biol Sci. 266 (1999) 1933-1938.

[11] E. Bullmore, O. Sporns, Complex brain networks: graph theoretical analysis of structural and functional systems, Nat. Rev. Neurosc. 10 (2009) 186-198.

[12] D.S. Callaway, J.E. Hopcroft, J.M. Kleinberg, M.E.J. Newman, S.H. Strogatz, Are randomly grown graphs really random?, Physical Review E 64 (2001) 041902.

[13] C. Castellano, M. Marsili, A. Vespignani, Nonequilibrium Phase Transition in a Model for Social Influence, Phys. Rev. E 66 (2002) 047104.

[14] V. Černý, Thermodynamical approach to the traveling salesman problem: An efficient simulation algorithm, Journal of Optimization Theory and Applications 45 (1985) 41-51.

[15] N.A. Christakis, J.H. Fowler, Social Network Sensors for Early Detection of Contagious Outbreaks, PLoS ONE 5(9) (2010) 12948.

[16] P. Clifford, A. Sudbury, A model for spatial conflict, Biometrika 60 (1973) 581-588.

[17] P. Csermely, Creative elements: network-based predictions of active centres in proteins and cellular and social networks, Trends in Biochemical Sciences 33(12) (2008) 569-578.

[18] S.N. Dorogovtsev, J.F.F. Mendes, Scaling properties of scale-free evolving networks: Continuous approach, Phys. Rev. E 63 (2001) 056125.

[19] R. Dunbar, Coevolution of neocortex size, group size and language in humans, Behavioural and Brain Sciences 16(4) (1993) 681–735.

[20] P. Erdős, A. Rényi, The Evolution of Random Graphs, Magyar Tud. Akad. Mat. Kutató Int. Közl. 5, (1960) 17–61.

[21] C.E. Gounaris, K. Rajendran, I.G. Kevrekidis, C.A. Floudas, Generation of Networks with prescribed degree-dependent clustering, Optimization Letters 5(3) (2011) 435-451.

[22] M.S. Granovetter, The Strength of Weak Ties, Am. J. of Sociol. 78(6) (1973) 1360-1380.

[23] X. Guardiola, A. Díaz-Guilera, C.J. Pérez, A. Arenas, M. Llas, Modeling diffusion of innovations in a social network, Phys. Rev. E 66 (2002) 026121.

[24] Q. Guo, T. Zhou, J. Liu, W. Bai, B. Wang, M. Zhao, Growing scale-free small-world networks with tunable assortative coefficient, Physica A 371 (2006) 814-822.

[25] D. Helbing, P. Molnár, Social force model for pedestrian dynamics, Phys. Rev. E 51 (1995) 4282.

[26] D. Helbing, I. Farkas, T. Vicsek, Simulating dynamical features of escape panic, Nature 407 (2000) 487-490.

[27] R.A. Holley, T.M. Liggett, Ergodic Theorems for Weakly Interacting Infinite Systems and the Voter Model, The Annals of Probability 3(4) (1975) 643-663.

[28] P. Holme, B.J. Kim, Growing scale-free networks with tunable clustering, Phys. Rev. E 65 (2002) 026107.

[29] R.L. Hughes, The flow of large crowds of pedestrians, Mathematics and Computers in Simulation 53 (2000) 367-370.

[30] A. Ianni, V. Corradi, The dynamics of public opinion under majority rules, Review of Economic Design 3 (2001) 257-277.

[31] M.J. Keeling, K.T. Eames, Networks and epidemic models, J. R. Soc. Interface 2 (2005) 295–307.

[32] I.G. Kevrekidis, C.W. Gear, J.M. Hyman, P.G. Kevrekidis, O. Runborg, C. Theodoropoulos, Equation-free coarse-grained multiscale computation: enabling microscopic simulators to perform system-level tasks, Comm. Math. Sciences 1(4) (2003) 715-762.

[33] B.J. Kim, Performance of networks of artificial neurons: The role of clustering, Phys. Rev. E 69 (2004) 045101.

[34] S. Kirkpatrick, C.D. Gelatt, M.P. Vecchi, Optimization by Simulated Annealing, Science 220 (1983) 671-680.

[35] K. Klemm, V.M. Eguiluz, Growing scale-free networks with small-world behavior, Phys. Rev. E 65 (2002) 057102.

[36] R. Kozma, M. Puljic, P. Balister, B. Bollobás, W.J. Freeman, Phase Transitions in the Neuropercolation Model of Neural Populations with Mixed Local and Non-Local Interactions, Biol. Cybern. 92 (2005) 367-374.

[37] R. Lambiotte, M. Ausloos, J.A. Hołyst, Majority model on a network with communities, Phys. Rev. E 75 (2007) 030101(R).

[38] C.C. Leary, M. Schwehm, M. Eichner, H.P. Duerr, Tuning degree distributions: Departing from scale-free networks, Physica A 382 (2007) 731–738.

[39] M. Llas, P.M. Gleiser, J.M. López, A.D. Guilera, Nonequilibrium phase transition in a model for the propagation of innovations among economic agents, Phys. Rev. E 68 (2003) 06610.

[40] V. Loreto, L. Steels, Social dynamics: Emergence of language, Nat. Phys. 3 (2007) 758 - 760.

[41] S. Maslov, K. Sneppen, Specificity and Stability in Topology of Protein Networks, Science 296(5569) 910-913.

[42] S. Mei, P. Sloot, R. Quax, Y. Zhu, W. Wang, Complex agent networks explaining the HIV epidemic among homosexual men in Amsterdam, Mathematics and Computers in Simulation 80(5) (2010) 1010-1030.

[43] N. Metropolis, A.W. Rosenbluth, M.N. Rosenbluth, A.H. Teller, E. Teller, Equation of State Calculations by Fast Computing Machines, J. Chemical Physics 21 (1953) 1087-1092.

[44] S. Milgram, The Small-World Problem, Psychology Today 1, (1967) 62-67.

[45] M.E.J. Newman, D.J. Watts, Renormalization group analysis of the small-world network model, Physics Letters A 263 (1999) 341-346.

[46] M.E.J. Newman, The structure and function of complex networks, SIAM Rev. 45 (2003) 167–256.

[47] J.M. Pacheco, F.L. Pinheiro, F.C. Santos, Population Structure Induces a Symmetry Breaking Favoring the Emergence of Cooperation, PLoS Comput Biol 5(12) (2009) 1000596.

[48] D.R. Parisi, C.O. Dorso, Microscopic dynamics of pedestrian evacuation, Physica A 354 (2005) 606-618.

[49] M. Salathé, M. Kazandjieva, J.W. Lee, P. Levis, M.W. Feldman, J.H. Jones, A high-resolution human contact network for infectious disease transmission, PNAS 107 (51) (2010) 22020-22025.

[50] M.A. Serrano, M. Boguna, Tuning clustering in random networks with arbitrary degree distributions, Phys. Rev. E 72 (2005) 036133.

[51] K.G. Spiliotis, C.I. Siettos, Computations on Neural Networks: from the Individual Neuron Interactions to the Macroscopic-level Analysis, Int. J. Bifurcation & Chaos 20 (2010) 121-134.

[52] K.G. Spiliotis, C.I. Siettos, A timestepper-based approach for the coarse-grained analysis of microscopic neuronal simulators on networks: Bifurcation and rare-events micro- to macro-computations, Neurocomputing 74(17) (2011) 3576-3589.

[53] L. Steels, J. Baillie, Shared grounding of event descriptions by autonomous robots, Robotics and Autonomous Systems 43 (2003) 163-173.

[54] S.H. Strogatz, Exploring complex networks, Nature 410 (2001) 268-276.

[55] C.J Tessone, System size stochastic resonance in a model for opinion formation, Physica A: Statistical Mechanics and its Applications 351 (2005) 106-116.

[56] E. Volz, Random networks with tunable degree distribution and clustering, Phys. Rev. E 70 (2004) 05611.

[57] J. Watts, S.H. Strogatz, Collective dynamics of 'small-world' networks, Nature 393 (1998) 440–442.

[58] F. Wu, A.B. Huberman, Novelty and collective attention, PNAS 104(45) (2007) 17599-17601.

**Figures**

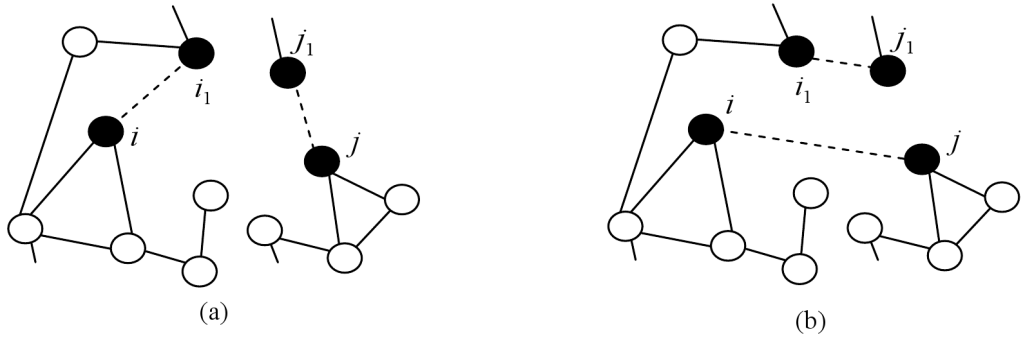

(a)          (b)

Figure 1.The proposed algorithm for tuning the APL of a network at will: (a) Select two nodes $i$ and $j$, having no mutual neighbors, each of whom having at least a neighbor, say $i_1$ and $j_1$ respectively, which do not form any triangle with their other neighbors or they have any mutual neighbors. (b) Select the edges that connect the nodes $i \Leftrightarrow i_1$ and $j \Leftrightarrow j_1$ and rewire them to connect $i \Leftrightarrow j$ and $i_i \Leftrightarrow j_1$.

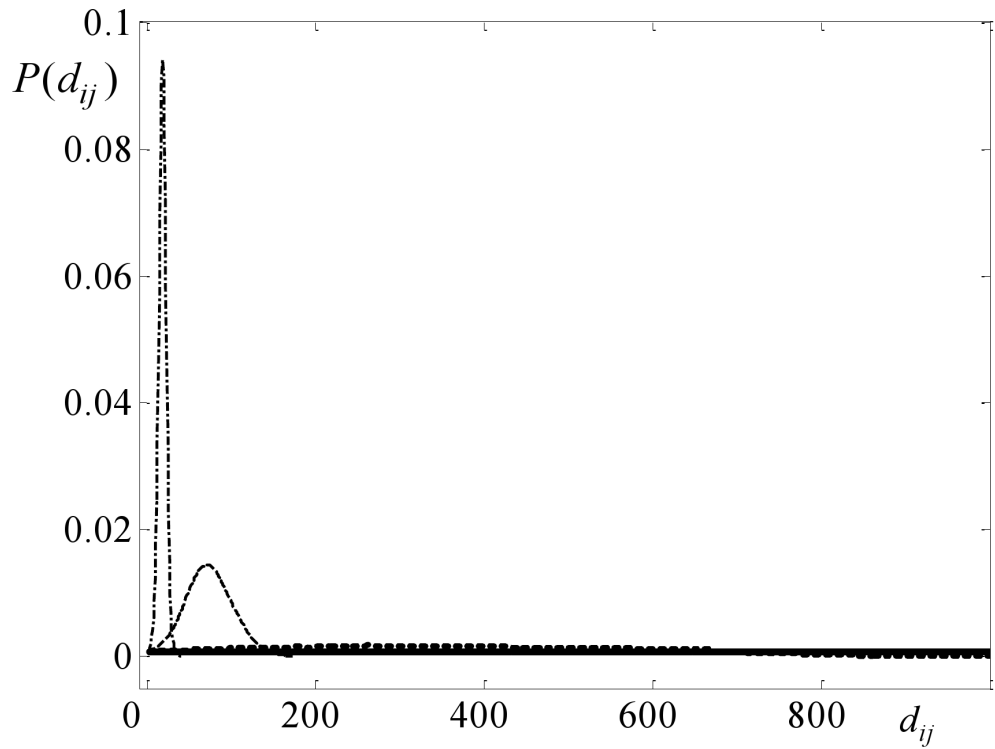

Figure 2. The probability distribution of the distance $d_{ij}$ between two nodes $i$ and $j$ of a network with $N = 10000$ nodes, $2k = 8$ when $p = 0$ (flat distribution, solid line), $p = 10^{-4}$ (dotted line), $p = 10^{-3}$ (dashed line) and $p = 10^{-2}$ (dashed-dotted line). The rewiring of few edges creates "long-range" connections resulting in more narrow-peaked distributions as the probability grows.

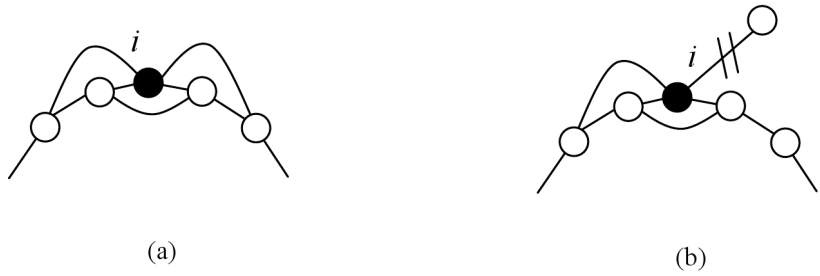

(a)                          (b)

Figure 3. (a) The connections of a node, $i$, in the case of the ring, when the probability is $p=0$. In this case, the proposed algorithm cannot find any possible edges to rewire, thus it cannot be implemented. (b) The connections of a node, $i$, in the case of the WS algorithm when $p>0$. The WS algorithm destroys the initial lattice formation and creates edges which can be selected by the proposed algorithm.

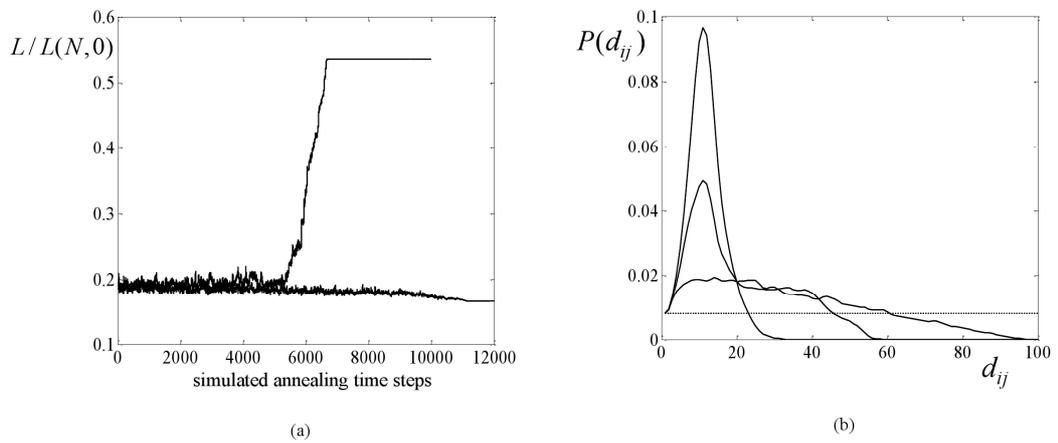

(a)                          (b)

Figure 4. (a) Implementation of the proposed algorithm to a network with $N=1000$ and $2k=8$, constructed initially with the WS algorithm and probability $p=0.01$ of rewiring edges. The APL reaches a maximum and minimum value. (b) Evolution of the probability distribution of the shortest distance between two nodes, comparing also with the uniform distribution of the ring network (dotted line) in the case of increasing the APL of the initial network. The, initially, peaked distribution changes to a more uniform one.

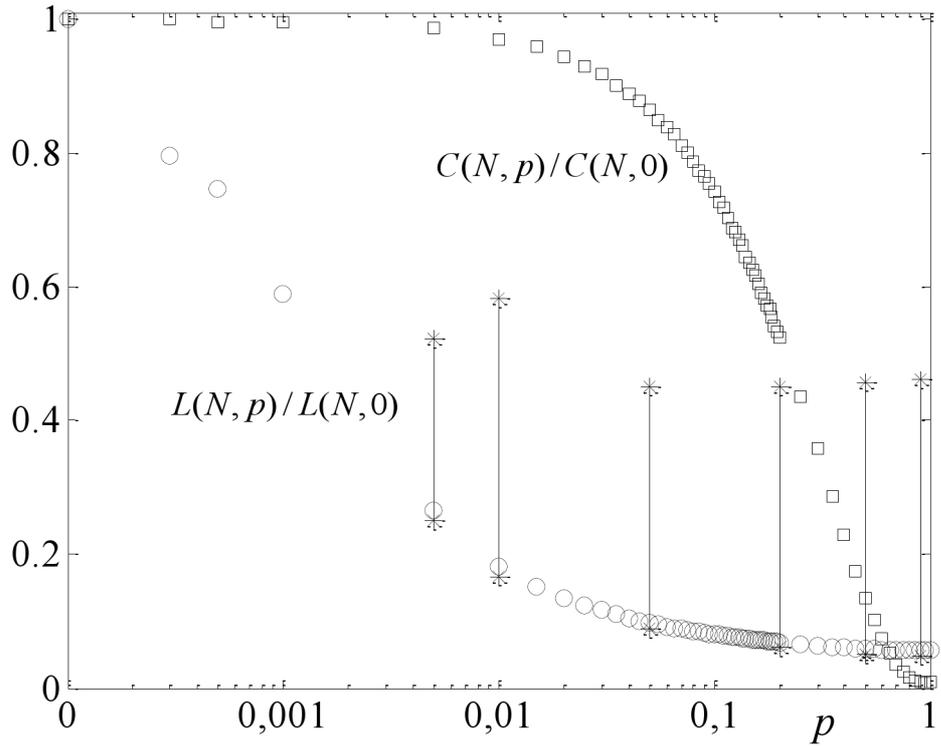

Figure 5. APL of "small-world" networks with $N = 1000$ and $2k = 6$ (depicted with circles), Clustering Coefficient (depicted with squares) and the maximum and minimum values (depicted with stars) of the APL that the proposed algorithm reached for networks constructed with probabilities $p = \{0.005, 0.01, 0.05, 0.2, 0.5, 0.9\}$ of rewiring edges. The clustering coefficients of the initially constructed networks do not change.

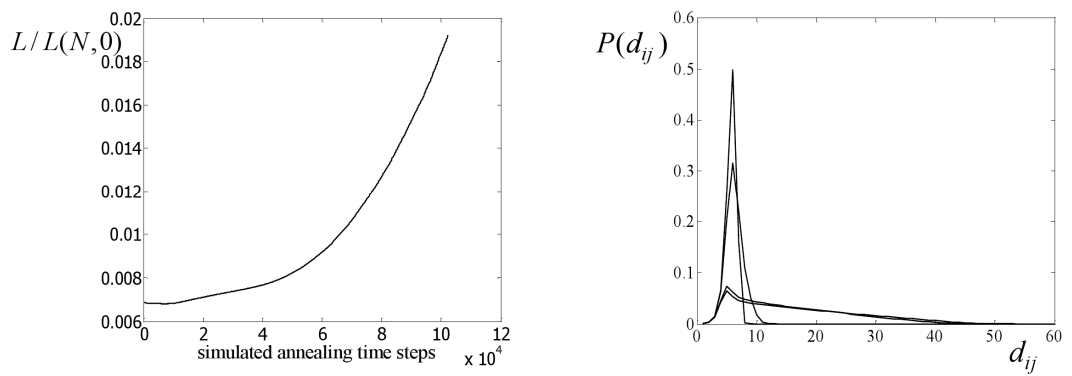

Figure 6. (a) Implementation of the proposed algorithm in the case of a network with $N = 10000$ nodes, $2k = 6$ initially constructed with a probability $p = 0.5$ of rewiring edges. (b) Evolution of the probability distribution of the shortest distance between two nodes as the Average Path Length increases.

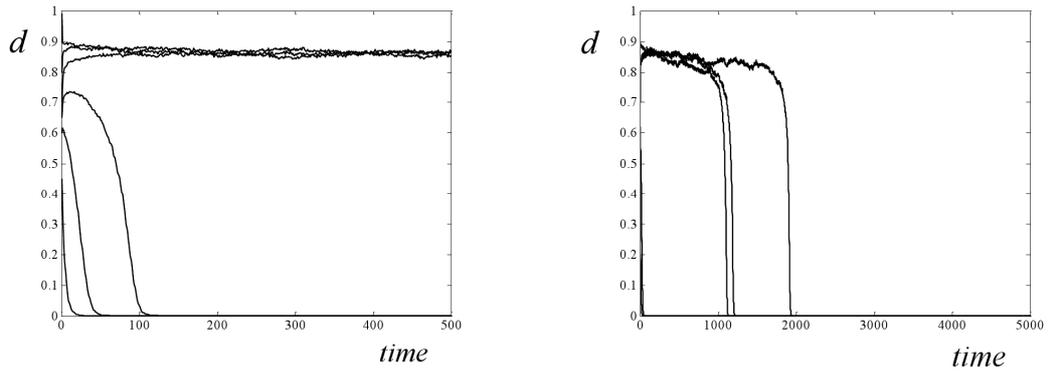

Figure 7. (a) Time evolution of the density of activated individuals, starting from different initial conditions, in the case of a small-world network initially constructed with a probability $p = 0.25$ of rewiring edges. (b) Time evolution of the density of activated individuals, starting from different initial conditions, in the case of a small-world network, initially constructed with a probability $p = 0.25$, in which the APL has been increased with the implementation of the proposed algorithm.

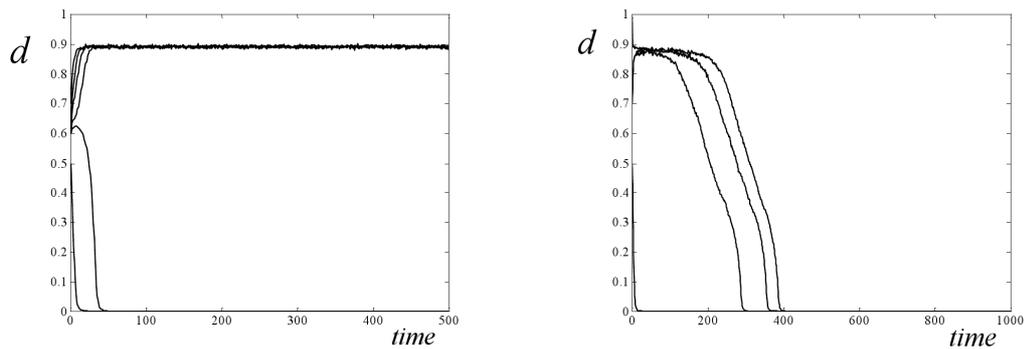

Figure 8. (a) Time evolution of the density of activated individuals starting from different initial conditions, in the case of a small-world network initially constructed with a probability $p = 0.7$ of rewiring edges. (b) Time evolution of the density of activated individuals, starting from different initial conditions, in the case of a small-world network, initially constructed with a probability $p = 0.7$, in which the APL has been increased with the implementation of the proposed algorithm.

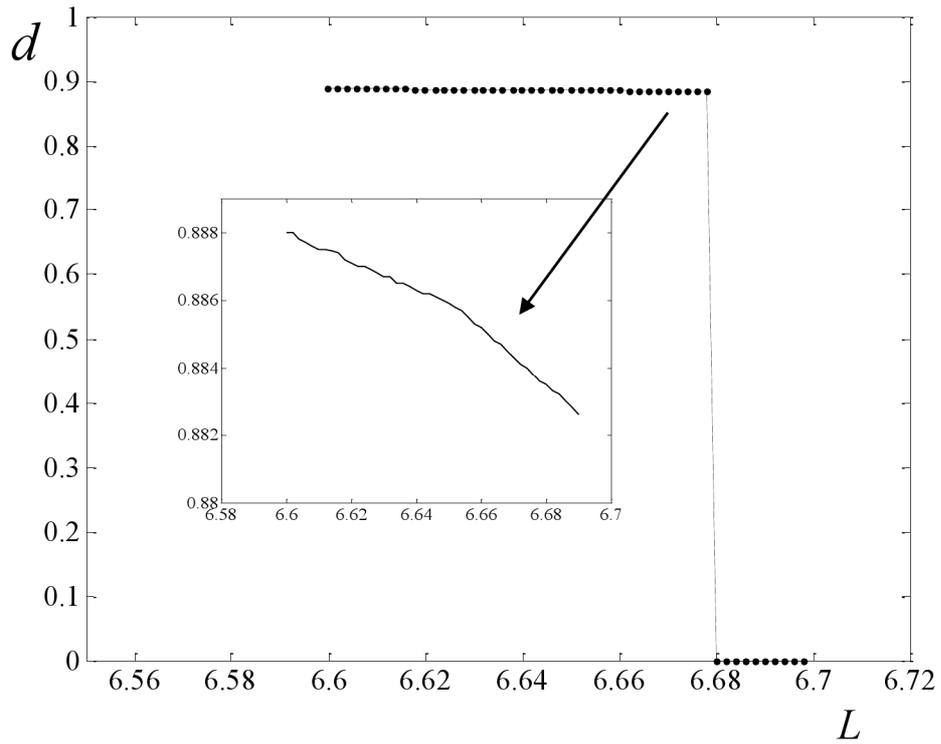

Figure 9. The coarse-grained bifurcation diagram of the density of activated individuals ($d$), with respect to the APL ($L$), in the case of the small-world network initially constructed with a probability $p = 0.25$ of rewiring edges.

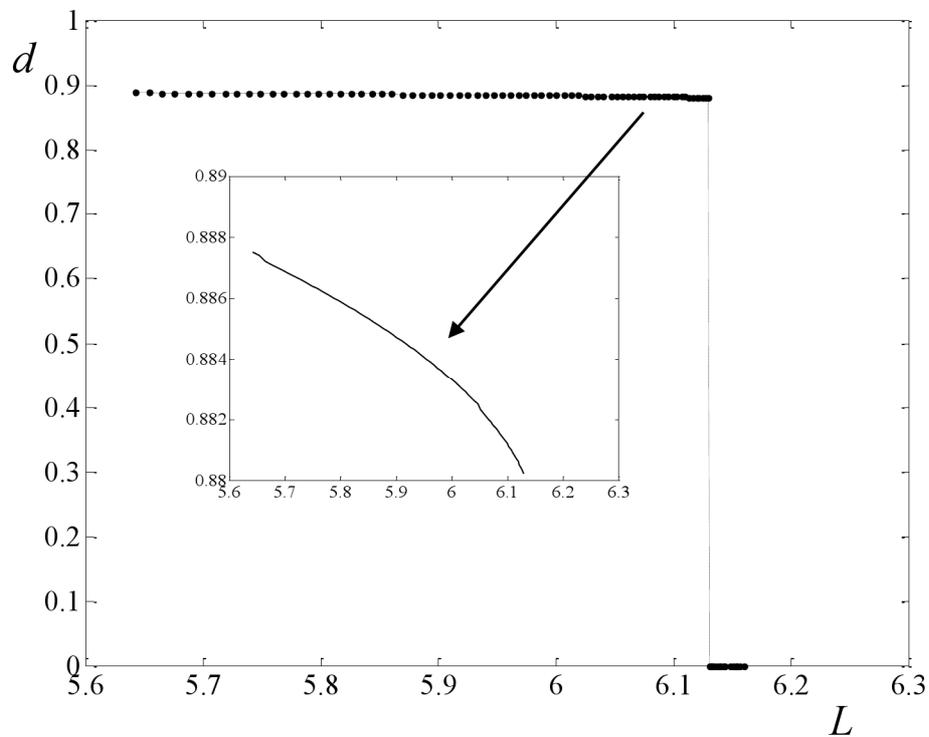

Figure 10. The coarse-grained bifurcation diagram of the density of activated individuals ($d$), with respect to the APL ($L$), in the case of the small-world network initially constructed with a probability $p = 0.7$ of rewiring edges.